# Optically controlled waveplate at a telecom wavelength using a ladder transition in Rb atoms for all-optical switching and high speed Stokesmetric Imaging

Subramanian Krishnamurthy[1], Y. Tu[1], Y. Wang[1], S. Tseng[1], and M.S. Shahriar[1,2,*]

[1]*Department of EECS, Northwestern University, Evanston, IL 60208, USA*
[2]*Department of Physics and Astronomy, Northwestern University, Evanston, IL 60208, USA*
*[\*shahriar@northwestern.edu](mailto:shahriar@northwestern.edu)*

**Abstract:** We demonstrate an optically controlled waveplate at ~1323 nm using the $5S_{1/2}$-$5P_{1/2}$-$6S_{1/2}$ ladder transition in a Rb vapor cell. The lower leg of the transitions represents the control beam, while the upper leg represents the signal beam. We show that we can place the signal beam in any arbitrary polarization state with a suitable choice of polarization of the control beam. Specifically, we demonstrate a differential phase retardance of ~180 degrees between the two circularly polarized components of a linearly polarized signal beam. We also demonstrate that the system can act as a Quarter Wave plate. The optical activity responsible for the phase retardation process is explained in terms of selection rules involving the Zeeman sublevels. As such, the system can be used to realize a fast Stokesmetric Imaging system with a speed of nearly 5 MHz. When implemented using a tapered nano fiber embedded in a vapor cell, this system can be used to realize an ultra-low power all-optical switch as well as a Quantum Zeno Effect based all-optical logic gate by combining it with an optically controlled polarizer, previously demonstrated by us. We present numerical simulations of the system using a comprehensive model which incorporates all the relevant Zeeman sub-levels in the system, using a novel algorithm recently developed by us for efficient computation of the evolution of an arbitrary large scale quantum system.

**OCIS codes**: (020.4180) Multiphoton processes; (020.1670) Coherent optical effects

## 1. Introduction

All-optical switching is important for optical communication and quantum information processing [1-5]. We recently demonstrated a ladder type modulator and proposed a novel scheme (employing high pressure buffer gas to broaden the atomic transitions) for a high-speed modulator (tens of GHz) that can be used to modulate signals at 1323 nm, using a control beam at 795 nm [6]. In addition, we have also described experimental schemes to realize such modulators at ultra-low powers (~40 pW) using a tapered nano fiber (TNF) [7, 8]. However, these modulators cannot be used as directional switches, as required in an optical communication network. In this paper, we report on the realization of an optically controlled waveplate and describe how this can be used as an all-optical switch in the telecommunication band. In combination with an optically controlled polarizer, recently demonstrated by us [9], the system can also be used to produce a Quantum Zeno Effect (QZE) based all-optical logic gate.

The optically controlled waveplate also has potential applications in Stokesmetric Imaging (SI). It is well documented that in many situations of interest, features indiscernible via conventional imaging become highly resolved under SI [10-13]. In a typical SI scenario, a target is illuminated by fully or partially polarized light. The light scattered or reflected by the target is then analyzed using a Stokesmeter, which determines the magnitude of each of the four Stokes parameter components. Stokesmeters, in their simplest form, are comprised of a combination of polarizers and wave-plates with different orientations. A key problem with the existing SI systems is that the polarizers and waveplates cannot be turned on or off or reoriented rapidly. The free space version of the optically controlled waveplate that is described in this paper has the potential to operate at speeds of ~5 MHz and thus holds the promise of making very high speed SI practical.

Rest of the paper is organized as follows. In section 2, we describe an ideal schematic for an optically controlled waveplate, using only four levels. A discussion of the non-idealities in the system and a comprehensive model that includes all the relevant Zeeman sub-levels are presented in section 3. We describe the experimental set-up in section 4 and a Jones matrix analysis of the system is provided in section 5. Experimental results and numerical simulations are presented in section 6. In section 7, we discuss the future work that will be pursued. Finally, in section 8, we present our conclusions.

## 2. Schematic of optically controlled waveplate and all-optical switch

In an atomic system involving ladder type transitions, the presence of two different frequencies open up the possibility of controlling the behavior of the probe (upper leg) polarization by careful design of the pump parameters (lower leg). In particular, it is possible to make the vapor cell act as a waveplate. The mechanism for producing controlled polarization rotation, is illustrated schematically in Fig. 1, using a cascaded atomic transition involving four levels where direct excitation to the upper level from the ground state is dipole forbidden. We consider the $m_F=0$ Zeeman sublevel of a certain hyperfine level into which the atoms have been optically pumped, as the ground state. The lower leg is excited by a right-circularly polarized ($\sigma_+$) beam (control beam), tuned a few GHz below the transition frequency to the intermediate level. The control beam, therefore, produces an off-resonant excitations to only the $m_F=1$ Zeeman sublevel in the intermediate state. The signal beam (probe), applied between the intermediate level and the upper level is chosen to be linearly polarized and hence has two components: $\sigma_+$ and $\sigma_-$. The $\sigma_-$ component sees the effect of the atoms and, because of the detuning, it sees only a real susceptibility with virtually no absorption, while the $\sigma_+$ component of the probe does not see any effect of the pump. The parameters of the control beam can be tuned to achieve the condition for a $\pi$ phase-shift for the $\sigma_-$ component only, so that at the output the linear polarization is rotated by 90 degrees.

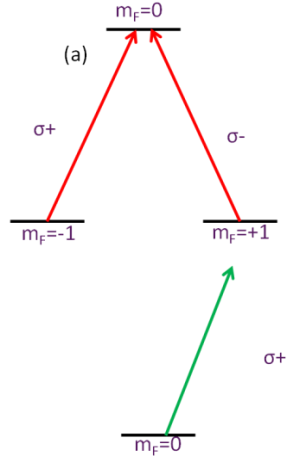

Fig.1. Schematic illustration of an optical switch using an optically controlled waveplate. See text for details.

It should be noted that the above schematic can only be used as a polarization rotator for a linear polarization of the probe. For example, it can be seen that it is impossible to obtain a purely circularl polarization of the probe, as it would require complete absorption on one of the legs of the upper transition. In fact, a simple Jones matrix analysis of the above system (in the absence of absorption) shows that the polarization state of the probe is $sin\ \phi\ \hat{x} + cos\ \phi\ \hat{y}$, where $\phi$ is the phase difference introduced between the two legs of the signal beam. In order for the system to behave as a quarter wave plate, for example, one needs to apply a linearly polarized pump. As shown in Fig. 2, in this scenario, both the $m_F = +1$ ($|2>$) and $m_F = -1$ ($|3>$) states get coupled to the $m_F = 0$ ($|1>$) state and the effect produced is not transparent in the simple schematic discussed above. However, by a suitable rotation of the intermediate states, one can reduce the process to a similar scenario, where for any arbitrary polarization of the pump, only one of the intermediate states is coupled, while the other is decoupled. Explicitly, any arbitrary polarization of the pump can be represented as $\alpha\hat{\sigma}_+ + \beta\hat{\sigma}_-$, where $\alpha$ and $\beta$ can be complex. This couples $|1>$ to $\alpha^*|2>+\beta^*|3>$ (denoted as $|+>$), where '*' denotes complex conjugation. The orthogonal state, $|-> = \alpha|3>-\beta|2>$, is not coupled by the pump, as can be verified by explicit calculation. A similar analysis can be carried out for the probe field as shown in Fig. 2. Now the situation is formally identical to the one presented in Fig. 1, where one of the legs sees the effect of the pump while the other does not.

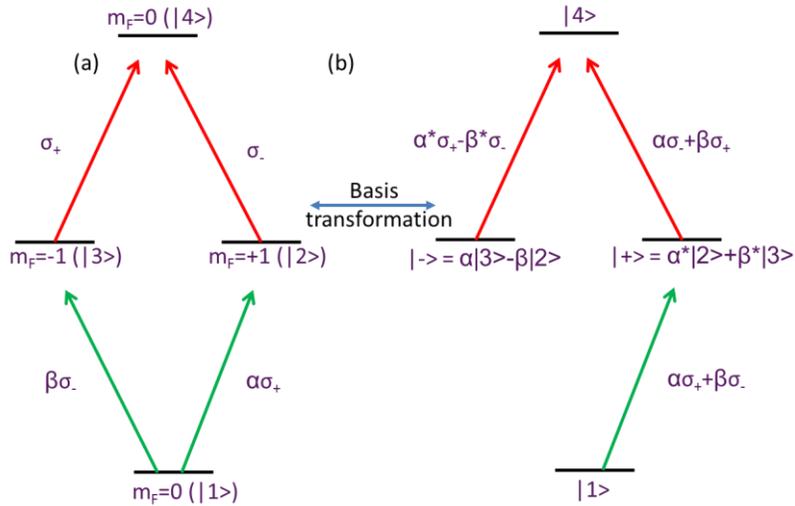

For illustrative purposes, let us consider a probe polarized along the $\hat{y}$ direction. Denoting $\hat{\sigma}_- = \hat{x} - i\hat{y}/\sqrt{2}$ and $\hat{\sigma}_+ = -\hat{x} - i\hat{y}/\sqrt{2}$, one obtains $\hat{y} = i(\hat{\sigma}_- + \hat{\sigma}_+)/\sqrt{2}$. Decomposing $\hat{y}$ in terms of the rotated basis and introducing a phase shift $\phi$ on the field coupling the $|+>$ state to $|4>$, one obtains the following expression for the polarization state ($\hat{p}$) of the probe

$$\hat{p} = i\,[(\alpha^* + \beta^*)(\alpha\hat{\sigma}_- + \beta\hat{\sigma}_+)e^{i\phi} + (\alpha - \beta)(\alpha^*\hat{\sigma}_+ + \beta^*\hat{\sigma}_-)]/\sqrt{2}$$

$$= i\left[\left((|\alpha|^2 + \beta^*\alpha)e^{i\phi} + (|\beta|^2 - \beta^*\alpha)\right)\hat{\sigma}_- + \left((|\beta|^2 + \alpha^*\beta)e^{i\phi} + (|\alpha|^2 - \alpha^*\beta)\right)\hat{\sigma}_+\right]/\sqrt{2} \qquad (1)$$

Using appropriate values of α,β and ϕ, one can obtain circular polarization. For example, consider a pump polarized linearly at $45^0$, for which α=i/(i-1) and β=1/(i-1). If we now set ϕ=π/2, one obtains $\hat{\sigma}_+$ polarization for the probe, as can be verified by explicit substitution in equation 1. In its most general form, the probe can also be in some arbitrary polarization state. All one needs to do is to resolve it in terms of the new rotated orthogonal states and introduce a phase shift on one of the legs, leaving the other unaffected. By a suitable choice of the pump polarization, one can place the probe in any desired polarization state.

In our set-up, we utilized the $5S_{1/2}$-$5P_{1/2}$-$6S_{1/2}$ cascade system in $^{87}$Rb atoms, with F=2 as the ground state. The pump and probe beams are at 795 nm and 1323nm respectively. Theoretical and experimental investigations of an optically controlled waveplate using ladder transitions in Rb have been carried out previously [14, 15]. However, both of these works employ the EIT effect where the upper leg is excited by a strong control field while the lower leg is probed by a weak optical field, and thus has fundamentally different characteristics than the system we have considered. Of course, the primary reason for choosing the upper leg as the probe is the need for an all-optical switch at a telecommunication wavelength.

In Fig. 1 and Fig.2, we showed a simplified set of energy levels in order to explain the basic process behind an optically controlled waveplate. In practice, however, it is extremely difficult to realize such an ideal system. We first note that it is virtually impossible to optically pump all the atoms into the , $5S_{1/2}$, F=2, $m_F = 0$ Zeeman sublevel. Hence, Zeeman sub-levels other than $m_F$=±1 at the $5P_{1/2}$ manifold also get coupled with the pump and probe optical fields. Furthermore, it is generally necessary to take into account both hyperfine levels (F'=1 and F'=2) in the $5P_{1/2}$ manifold to account for Doppler broadening and power broadening. Thus, the full set of energy levels that need to be considered is quite large, and the actual model employed for our system is discussed in the next section. For the remainder of the paper, the hyperfine levels in the ground state are indicated by unprimed alphabets (F), those in the $5P_{1/2}$ level are primed (F') and those in the $6S_{1/2}$ level are double-primed (F").

## 3. Comprehensive model used for numerical simulation

In previous analyses of similar systems, a simple model consisting of only the relevant hyperfine levels transitions was employed [14, 15]. Other works [16] consider some of the Zeeman sub-levels, but make use of few assumptions to eliminate some of the density matrix elements to arrive at a somewhat approximate result. In our model, we considered all the Zeeman sub-levels which explicitly interact with an optical field (all sub-levels of the F=2, F'=1,2 and F"=1 hyperfine levels), while the F=1 hyperfine level and the $5P_{3/2}$ level were only considered as population transfer levels and hence all their sub-levels were lumped together as a single level. The full set of energy levels that we have incorporated in our model are shown in Fig. 2. The transition strengths [17] indicated are expressed as multiples of the weakest transition, which in our case is the transition from the F=2, $m_F$=0 sub-level to the F'=1, $m_F$=1 sub-level, for example. We assume that the control beam is tuned below the F=2→ F'=1 transition while the signal beam is detuned by an amount $\delta_s$ from the F'=1→ F"=1 transition. Due to the Doppler width and power broadening, the F'=2 hyperfine level also interacts with both the control and the signal optical fields (indicated by dashed lines), albeit at a large detuning, and these interactions have been taken into account in our model. However, we ignored the coherent coupling between F=1 and the $5P_{1/2}$ manifold, because of the large frequency difference between F=1 and F=2 (~6.8GHz for $^{87}$Rb).

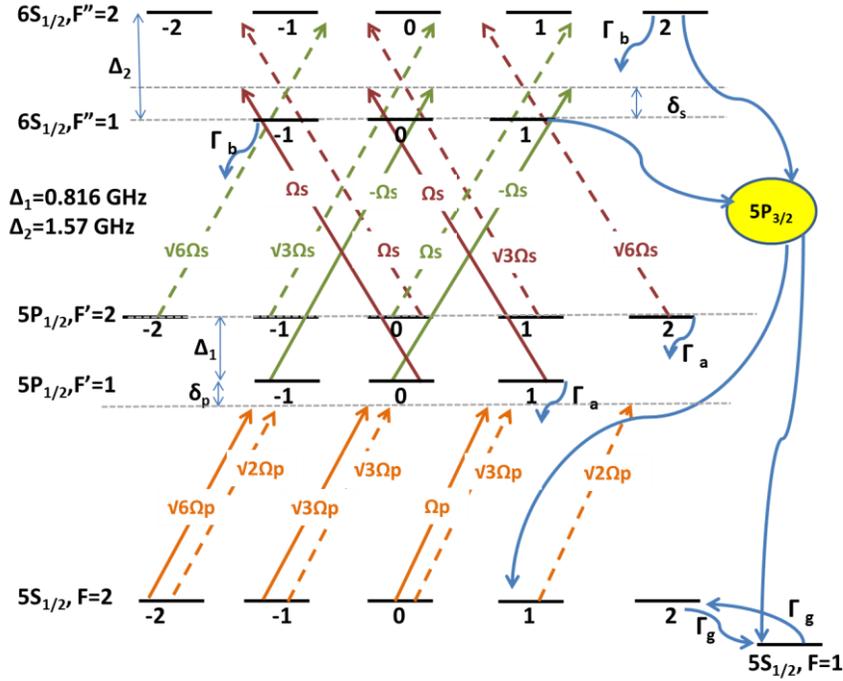

Fig. 2 Model used for numerical computation. See text for more details.

All the Zeeman sub-levels in the $5P_{1/2}$ ($6S_{1/2}$) manifold are assumed to decay at the same rate, $\gamma_a$~5.75 MHz ($\gamma_b$~3.45 MHz). We also assume a nominal cross-relaxation rate ($\gamma_g$ ~0.01 MHz) between the ground states. The decay rate between any two Zeeman sub-levels was calculated by using the fact that it is proportional to the square of the matrix element of the corresponding transition, and that the sum of all such decays rates from the decaying level must equal the net decay from that level. For example, consider $m_F$=0, F'=2 sub-level which decays at a rate $\gamma_a$. The transition strengths for the $\sigma_+$, $\sigma_-$, and $\pi$-transitions to the F=2 (F=1) are in the ratio $\sqrt{3}$:$\sqrt{3}$:0 (1:1:2). Thus, the net decay rate between the $m_F$=0, F'=2 sub-level to the $m_F$=-1,+1 and 0 states of the F=2 level were computed to be $\gamma_a$/4, $\gamma_a$/4 and 0 respectively and the decay to the F=1 level was computed to be $\gamma_a$/2, since all the hyperfine levels in the F=1 state are lumped together as a single state in our model. We have also considered the sourcing of atoms into the ground states from the $6S_{1/2}$ manifold via the $5P_{3/2}$ state. A detailed calculation, taking into account the various branching ratios into and from all the hyperfine levels of the $5P_{3/2}$ state was used to determine these "effective decay rates" directly from the $6S_{1/2}$ states to the ground states. Table 1 shows these "effective decay rates" from each of the Zeeman sub-levels in the $6S_{1/2}$ manifold to the ground states. The ratio between the rates of decay into the $5P_{1/2}$ and $5P_{3/2}$ states from a given upper level was decided by the ratio of the explicit values of the transition strength of the D1 and D2 lines [17].

| | | F"=2 | | | | | F"=1 | | |
|---|---|---|---|---|---|---|---|---|---|
| | | $m_F$=-2 | $m_F$=-1 | $m_F$=0 | $m_F$=1 | $m_F$=2 | $m_F$=-1 | $m_F$=0 | $m_F$=1 |
| | $m_F$=-2 | 0.68852 | 0.19426 | 0.05055 | 0 | 0 | 0.2361 | 0.09722 | 0 |
| | $m_F$=-1 | 0.19426 | 0.47296 | 0.190277 | 0.07583 | 0 | 0.1667 | 0.11805 | 0.04861 |
| F=2 | $m_F$=0 | 0.05055 | 0.190277 | 0.45166 | 0.190277 | 0.05055 | 0.104167 | 0.125 | 0.104167 |
| | $m_F$=1 | 0 | 0.07583 | 0.190277 | 0.47296 | 0.19426 | 0.04861 | 0.11805 | 0.1667 |
| | $m_F$=2 | 0 | 0 | 0.05055 | 0.19426 | 0.68852 | 0 | 0.09722 | 0.2361 |
| | $m_F$=-1 | 0.04722 | 0.03333 | 0.02083 | 0.009722 | 0 | 0.21296 | 0.1226875 | 0.1088 |
| F=1 | $m_F$=0 | 0.01944 | 0.023611 | 0.025 | 0.023611 | 0.01944 | 0.1226875 | 0.199 | 0.1226875 |
| | $m_F$=1 | 0 | 0.009722 | 0.02083 | 0.03333 | 0.04722 | 0.1088 | 0.1226875 | 0.21296 |

Table 1. Effective Decay rates between excited states and ground states

We used the Liouville equation, which describes the evolution of the density matrix in terms of a commutator between the density matrix and the Hamiltonian augmented by the phenomenological determined decay rates, to obtain the steady-state solution. The usual method of vectorizing the density matrix and then inverting the coefficient matrix thus obtained, is not easy to handle as the size of the coefficient matrix is very large (400*400). In order to overcome this problem, we made use of a novel algorithm, recently developed by us [18] which would compute the said coefficient matrix automatically in a very efficient manner, given the Hamiltonian and the source matrix. While averaging over the Doppler profile, we used the supercomputing cluster at Northwestern (QUEST) to perform our computations. Using 64 cores and computing the steady state solution for 512 values of detuning, each averaged over 800 points of the Doppler profile, we obtained the steady-state solution for our 22-level system in 3-4 minutes.

## 4. Experimental set-up

The experimental configuration is illustrated schematically in Fig. 4. Briefly, beams from two tunable lasers (one at 795 nm, and the other at 1323 nm) are combined with a dichroic mirror (DCM). A part of the 795 light is sent to a reference vapor cell for saturated absorption spectroscopy and locking. The combined beams are sent through a vapor cell, shielded from magnetic fields with µ-metal. The cell is heated using bifiliarly wounded wires that do not add any magnetic fields. After passing through the cell, another DCM is used to split the light into two parts, and each frequency is detected with a separate detector. The control beam at 795 nm, initially polarized linearly, is passed through a quarter-wave-plate in order to produce circular polarization. The polarization of the signal beam, at 1323 nm, is controlled separately with a half-wave-plate. Ideally, the 1323 nm laser would also be locked at a particular frequency but this laser was found to be stable, so that locking it was not necessary.

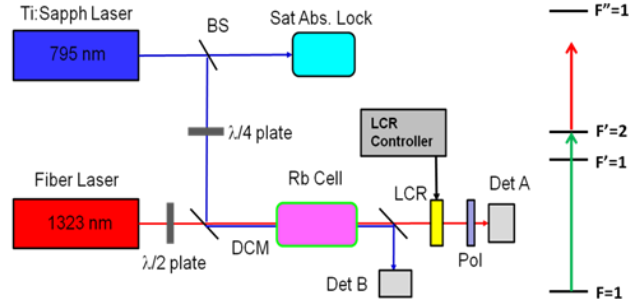

Fig. 4. Experimental Set-up

In order to analyze the polarization of the signal beam after passing through the cell, we inserted an analyzer before detector A, consisting of a voltage-controlled liquid crystal retarder (LCR), whose fast axis is placed at 45 degrees to the initial polarization direction (say $\hat{y}$) of the signal beam, followed by a polarizer with its axis orthogonal to initial polarization of the signal beam. Having a orthogonal polarizer at the output is, in general, not enough, as one cannot determine if the observed signal is a manifestation of polarization rotation alone or a combination of both rotation and absorption, unless of course the detector signal is at least as large as the far off-resonant signal. However, it can serve as a diagnostic tool in identifying the regions of large phase shift as the probe is scanned across the $6S_{1/2}$ manifold. The LCR produces a phase retardance between its orthogonal axes that depends non-linearly on the amount of voltage applied to the LCR controller which can determined from a calibration curve provided by the manufactured and verified independently by us. During our experiment, the control voltage to the LCR is scanned linearly from 0 V to 10 V, with 2 V and 8V corresponding to a phase shifts of approximately $\pi$ and 0, respectively for the wavelength that we are using. This particular arrangement of the analyzer provides us with a very large set of data points (corresponding to the LCR scan) from which to obtain the values of absorption coefficients and the phase rotations, thus making it more robust against noise in the system. The signal observed at the detector A can be ascertained by performing a Jones matrix analysis of the entire system, which is discussed next.

## 5. Jones matrix analysis

Let $E_{in}$ be the signal field amplitude before the Rb cell and let $\hat{y}$ be its polarization direction. Let $\hat{\sigma}_+ = -\left(\frac{\hat{x}+i\hat{y}}{\sqrt{2}}\right)$ and $\hat{\sigma}_- = \left(\frac{\hat{x}-i\hat{y}}{\sqrt{2}}\right)$ be the unit vectors corresponding to right (RCP) and left circular polarizations (LCP) respectively. Then, $E_{in}$ can be represented in the circular polarization basis as $\vec{E}_{in} = E_0\hat{y} = E_{in+}\hat{\sigma}_+ + E_{in-}\hat{\sigma}_-$, where $E_{in+} = E_{in-} = i E_0/\sqrt{2}$ and $E_0$ is some arbitrary value. The field amplitude after the cell can be represented, in its most general form as $\vec{E}_{after\ cell} = E_{in+}e^{(-\alpha_+ + j\phi_+)}\hat{\sigma}_+ + E_{in-}e^{(-\alpha_- + j\phi_-)}\hat{\sigma}_-$, where $(\alpha_+, \varphi_+)$ and $(\alpha_-, \varphi_-)$ are the attenuations and phase rotations for the RCP and LCP part of the signal beam respectively after passing through the

Rb vapor cell. With $\hat{x}$ and $\hat{y}$ as the basis for the Jones vector representation and after some algebraic manipulation, we find that the field amplitude after the cell can be represented as

$$E_{\text{after cell}} = \frac{iE_0}{2} e^{(-\alpha_- + j\phi_-)} \begin{bmatrix} -e^{(-\alpha_d + j\phi_d)} + 1 \\ -i\left(e^{(-\alpha_d + j\phi_d)} + 1\right) \end{bmatrix}, \text{ where } \begin{matrix} \alpha_d = \alpha_+ - \alpha_- \\ \phi_d = \phi_+ - \phi_- \end{matrix}$$

Thus, $\alpha_d$ and $\phi_d$ represent the differential absorption and phase rotation between the RCP and LCP parts of the signal beam. If $\theta$ represents the phase retardation produced by the LCR, then the Jones matrix for the LCR is given by

$$J_{LCR} = \begin{bmatrix} \exp(i\theta/2) & 0 \\ 0 & \exp(-i\theta/2) \end{bmatrix}$$

and the Jones matrix for the LCR whose axis is rotated by $45^0$ is given by $J_{LCR45}=R^{-1}(45^0)J_{LCR}R(45^0)$ where $R(45^0)$ represents the rotation matrix for $45^0$ and is given by

$$R(45) = \begin{bmatrix} \frac{1}{\sqrt{2}} & \frac{1}{\sqrt{2}} \\ \frac{-1}{\sqrt{2}} & \frac{1}{\sqrt{2}} \end{bmatrix}$$

Finally, the polarizer with its axis parallel to the $\hat{y}$ axis has the Jones matrix representation

$$J_{XPol} = \begin{bmatrix} 1 & 0 \\ 0 & 0 \end{bmatrix}$$

Thus, the Jones vector for the signal observed at the detector A would be

$$J_{out} = J_{YPol} * J_{LCR45} * E_{aftercell}$$

Performing the calculations, we find that the intensity as seen by the detector A is given by,

$$I = \frac{E_0}{4} e^{-2\alpha_-} (1 + e^{-2\alpha_d} + (1 - e^{-2\alpha_d})\sin\theta - 2e^{-\alpha_d}\cos\phi_d \cos\theta)$$

In Fig. 5(a), I is plotted for different values of $\phi_d$ (in degrees) and $\alpha_d=0$ as the LCR phase retardance $\theta$ varies from 0 (left-end) to $\pi$ (center) and back to 0 (right-end). As is evident from the figure, the signature for increasing differential phase rotation is the upward shift of the minimum and downward shift of the maximum of the curve until $\phi_d = 90^0$, at which point the signal is perfectly flat. For greater values of $\phi_d$, the shape of the curve gets inverted until $\phi_d = \pi$. On the other hand, for non-zero values of $\alpha_d$ and $\phi_d =0$, the minima of the curves get shifted inwards, the curves slope upward on either side of the minima and the central part of the curve is flattened out, as shown in Fig. 5(b). For non-zero values of both $\alpha_d$ and $\phi_d$, the interpretation is not so straight-forward and one has to use 3 data points and invert the expression for I to obtain their values. The algebra is somewhat involved and we present only the final result –

$$y = e^{\alpha_d} = \sqrt{\frac{I_2^2(C_1 - C_3 - S_{3-1}) + I_1 I_2(C_3 - C_2 - S_{2-3}) + I_2 I_3(C_2 - C_1 - S_{1-2})}{I_2^2(C_3 - C_1 - S_{3-1}) + I_1 I_2(C_2 - C_3 - S_{2-3}) + I_2 I_3(C_1 - C_2 - S_{1-2})}} \text{ and}$$

$$\cos\phi_d = \frac{(I_2 - I_1)\left(y + \frac{1}{y}\right) + (I_1 S_2 - I_2 S_1)\left(\frac{1}{y} - y\right)}{2(I_2 C_1 - I_1 C_2)},$$

Where $C_j = \cos(\theta_j)$, $S_j = \sin(\theta_j)$, and $S_{j-k} = \sin(\theta_j - \theta_k)$, where $\theta_j$ is the phase rotations produced for some voltage, and $I_j$ is the corresponding intensity seen by the detector.

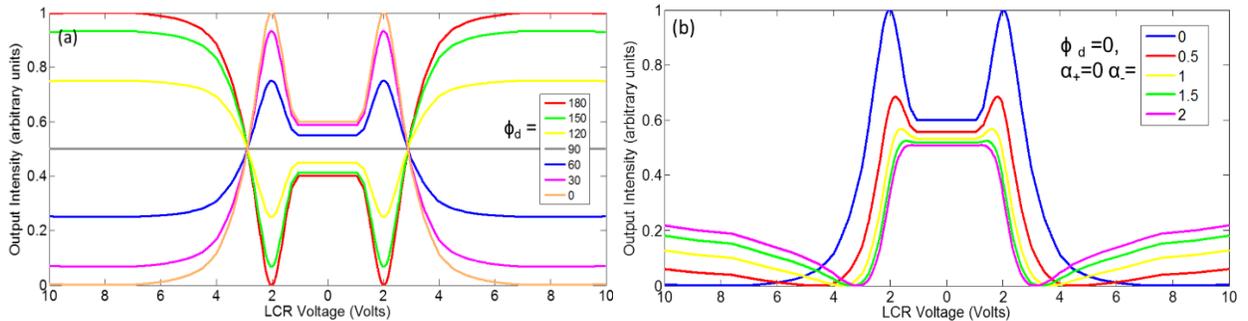

Fig. 5 Ideal Output seen by detector for different values of a) $\phi_d$ with $\alpha_+ = \alpha_- = 0$ and b) $\alpha_-$ with $\phi_d = \alpha_+ = 0$

## 6. Results

The exact spectroscopic details of the 5P$_{1/2}$-6S$_{1/2}$ transitions depend critically on the control beam intensity and detuning, the temperature of the cell and probe detuning. Thus, as a diagnostic tool to identify regions of high phase shift, an orthogonal polarizer was placed after the Rb cell. As the probe laser was scanned across the 6S$_{1/2}$ manifold, the detuning of the control beam was varied in order to maximize the transmission through the orthogonal polarizer over the largest possible bandwidth. Then, with the control beam and signal lasers positioned at the detunings previously determined, the LCR control voltage was scanned in order to estimate the differential phase shift and attenuation.

Figure 6(b) shows the transmission through an orthogonal polarizer as the probe laser was scanned over ~5 GHz. The control beam was right circularly polarized and the signal beam was vertically polarized. The pump was placed at a detuning of about 1.2 GHz. The temperature of the cell was maintained around 130º Celsius and pump power was about 600 mW, obtained from a Ti-Sapphire laser. The probe laser was about 1 mW obtained from a fiber coupled semiconductor laser. Both beams were focused to a spot size of about 50 μm near the center of the Rb cell. The frequency scan was then stopped and with the probe laser fixed at the regions of high phase shift, the LCR control voltage was scanned linearly from 10V to 0V and back up to 10V. The blue trace (normalized from 0 to 1) corresponds to the situation when the control beam was blocked and can thus be treated as the reference signal, corresponding to 0 phase retardance. When the control beam is unblocked, our system acts as an optically controlled waveplate and the red trace is obtained. Comparing with the theoretical plots, one can see that the phase shifts in Fig. 6(a) and Fig. 6(c) correspond to $\phi_d \sim 160^0$ and $\phi_d \sim 180^0$, respectively with $\alpha_d \sim 0$. The values obtained are consistent with those obtained using the analytical expressions.

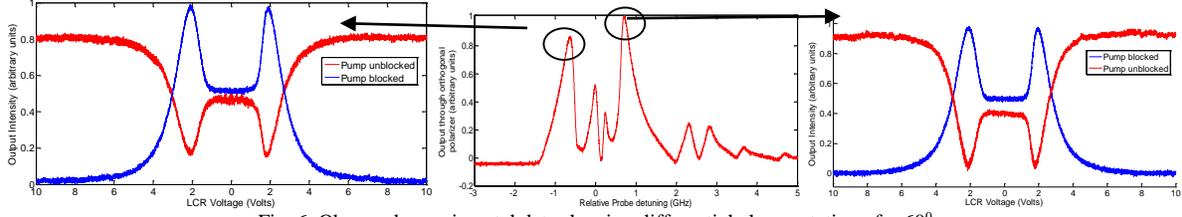

Fig. 6. Observed experimental data showing differential phase rotation of ~ 60$^0$ and almost no differential absorption. See text for more details

Fig. 7 shows the result obtained from numerical simulations using the model presented in section 2. We perform our calculations by setting $\Gamma_a$ to unity and rescaling all parameters in units of $\Gamma_a$. The pump is resonant with the F=1 to F'=2 transition and the probe detuning ($\delta_s$) ranges from -1200$\Gamma_a$ to 1200$\Gamma_a$. Fig. 7(a) and 7(b) show the phase shift of the RCP and LCP parts of the signal beam introduced by the Rb medium and Fig. 7(c) shows the difference between them. Fig. 7(d)-7(f) show the corresponding figures for attenuation. As is evident from the figure, at a pump detuning of ~1.2 GHz and for Rb density of $10^{12}$/cm$^3$ and cell length of 7.5 cm, we can produce a phase shift of about 180$^0$ with minimal differential absorption. The relevant parameters used for this particular simulation are as follows - the decay rates $\Gamma_a$, $\Gamma_b$ and $\Gamma_g$ are 2π*5.75 sec$^{-1}$, 2π*3.45 sec$^{-1}$ and 2π*0.1 sec$^{-1}$ respectively. The separation Δ, between F'=1 and F'=2 is 2π*814.5 sec$^{-1}$ (= 141.4$\Gamma_a$) and the Rabi frequencies have been chosen to be $\Omega_p = 100\Gamma_a$, and $\Omega_s = 0.1\Gamma_a$. The expressions used to calculate the attenuation and additional phase retardance introduced by the Rb medium are given by

Phase Shift:
$$\phi_+ = kL\frac{\beta_+}{2} Re(a_{13,4}\rho_{13,4} + a_{14,5}\rho_{14,5} + a_{12,7}\rho_{12,7} + a_{13,8}\rho_{13,8} + a_{14,9}\rho_{14,9})$$
$$\phi_- = kL\frac{\beta_-}{2} Re(a_{12,5}\rho_{12,5} + a_{13,6}\rho_{13,6} + a_{12,9}\rho_{12,9} + a_{13,10}\rho_{13,10} + a_{14,11}\rho_{14,11})$$

Attenuation:
$$\alpha_+ = kL\beta_+ Im(a_{13,4}\rho_{13,4} + a_{14,5}\rho_{14,5} + a_{12,7}\rho_{12,7} + a_{13,8}\rho_{13,8} + a_{14,9}\rho_{14,9})/2$$
$$\alpha_- = kL\beta_- Im(a_{12,5}\rho_{12,5} + a_{13,6}\rho_{13,6} + a_{12,9}\rho_{12,9} + a_{13,10}\rho_{13,10} + a_{14,11}\rho_{14,11})/2$$
And
$$\beta_\pm = b_{min}^2 \frac{3n_{atom}\Gamma\lambda^3}{4\pi^2\Omega_{min}},$$

where $k$ is the wavevector of signal beam, L is the length of the cell, $n_{atom}$ is the density of Rb atoms, $\Omega_{min}$ is the Rabi frequency for the weakest probe transition and the various $a_{ij}$'s are the ratios of the Rabi frequency ($\Omega_{ij}$) of the $|i\rangle$-$|j\rangle$ transition to $\Omega_{min}$. For example, $a_{12,7} = \Omega_{12,7}/\Omega_{14,9} = \sqrt{6}$. $b_{min}^2$ is the fraction of the atoms (<1) that decay along the transition corresponding to $\Omega_{min}$, among all allowed decay channels from the decaying level. In our model, the amplitudes for all possible transitions from $|14\rangle$ are in the ratio $1:1:1:\sqrt{3}:\sqrt{6}$ and hence the fraction of atoms that decay along the different channels are in the ratio 1:1:1:3:6. Thus, $b_{min}^2 = 1/(1+1+1+3+6) = 1/12$.

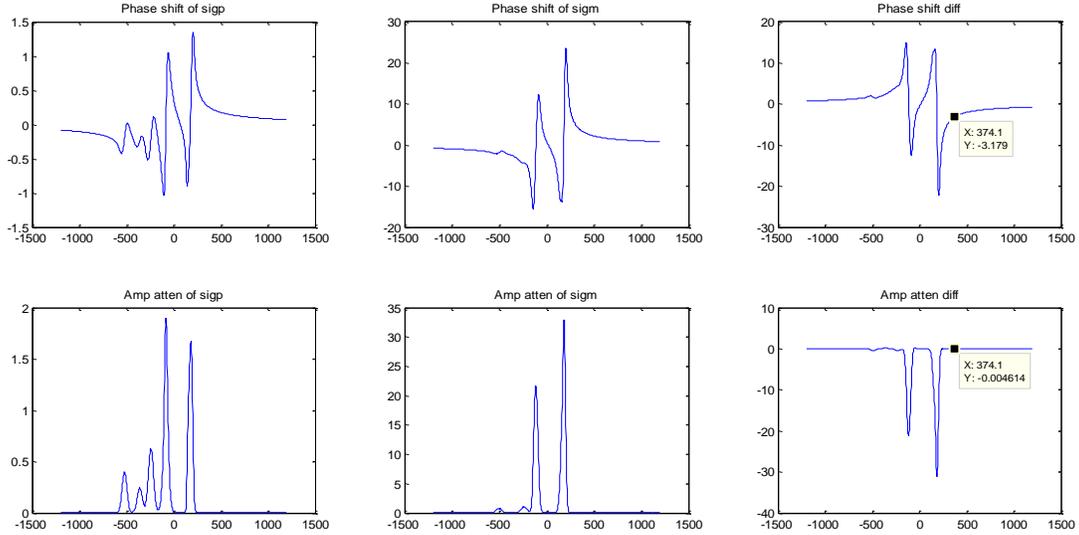

Fig. 7. Numerical simulation of 15-level system showing phase shift and attenuation of the RCP and LCP parts of the probe beam as a function of probe detuning. Here $\delta_c \sim 1.2$ GHz, $n_{atom} \sim 10^{12}$/cm$^3$ and $\Omega_{min}=100\Gamma_a$. See text for more details.

It is possible to produce arbitrarily large phase rotation by increasing the temperature and thus the density of Rb atoms. However, since the difference in phase rotation is so large, the bandwidth over which the device can operate as a half-wave plate, for example, becomes narrower. Thus, for wide bandwidth operation, we would need to operate in the parameter space where we not only have $\phi_d \sim 180^0$, but the slope should also be very small, as shown in Fig. 7(c). We also investigated the waveplate effect under a co-propagating geometry. We found that the absorption line shapes for the co-propagating geometry are broader and shallower compared to the counter-propagating geometry, the reasons for which have been clearly elucidated in Ref 9. As a consequence, under identical conditions, the phase rotation produced is smaller. This was confirmed experimentally, but the results have been omitted for the sake of brevity.

The system can also be used as a quarter-wave plate by using a pump that is linearly polarized at $45^0$ to the direction of polarization of a linearly polarized probe, as explained earlier [9]. Fig. 8a shows the result for such a situation with the output polarizer parallel to the direction of polarization of the probe. Here, the probe is vertically polarized and the pump is linearly polarized at $45^0$, is at a detuning of $\delta_c \sim 1$ GHz and has a power of ~600 mW. Fig. 8b shows the corresponding simulation results for a vertical and circularly polarized probe after passing through our analyzing system.

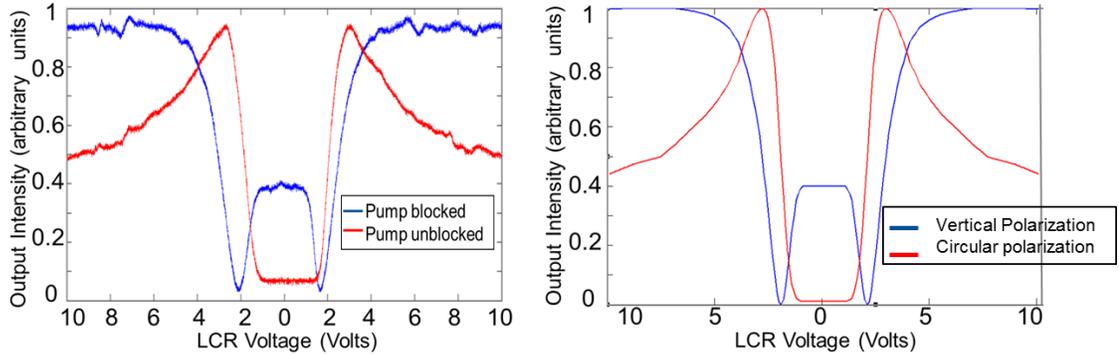

Fig 8. System behaving as Quarter Wave Plate using pump polarized at $45^0$

Finally, for certain frequencies, it was observed that that the absorption profile changed dramatically from 0 to approximately 80% when the polarization of the pump was changed by 3-4 degrees, by using a half-wave plate. Fig. 9 shows the result obtained for pump power ~ 600mW and a detuning of ~1 GHz. Such a device could be an inexpensive tool for detecting the polarization of an optical field with high degree of accuracy. Again, the exact location of such behavior depends critically on the density of Rb, the pump power and detuning and the temperature.

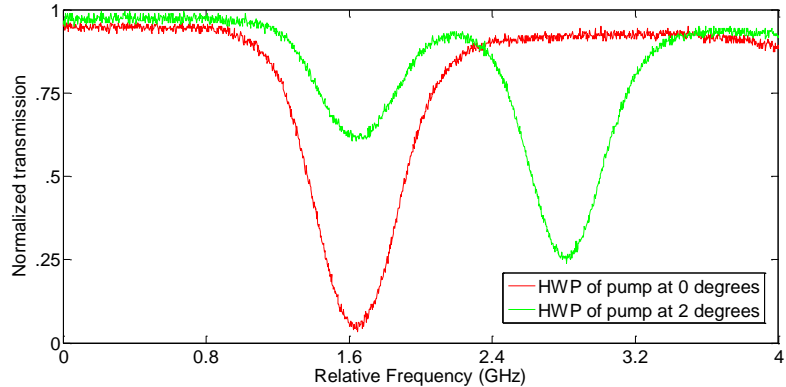

## 7. Future Work

In a closely related experiment, we have also demonstrated an optically controlled polarizer, where a control laser at 795 nm is used to realize a polarizer for the signal beam at 1323 nm. Once the optically controlled waveplate and polarizer are combined in the same cell, it should be possible to demonstrate a Quantum Zeno Effect (QZE) based all optical logic gate [9]. Such a gate, realized with a TNF embedded in vapor cell, is expected to require pump power as low as a few nanowatts and thus has potential applications in quantum information processing. In a TNF, the typical mode area is ~0.2 μm$^2$ [7,8]. Thus, assuming a saturation intensity of 3mW/cm$^2$, a Rabi frequency of $10\gamma_a$ would correspond to a power of only ~1 nW. Unlike the free space case, the maximum speed of operation in the TNF system would be limited by the transit time broadening (~60 MHz) rather than the natural linewidth of the 5P manifold (~6 MHz). To see why, note that the speed is limited by the rate at which atoms in the intermediate state relax to the ground state [6] and for a TNF system, this rate is effectively determined by the transit time. A more thorough investigation is needed to identify the parameter space for optimum operation of both the polarizer and waveplate effect simultaneously. We have indicated some of the ways to improve the performance of the optically controlled polarizer [6]. Significant changes to the set-up are required in order to implement the improved scheme both in free space and in the TNF system. Efforts are underway in our laboratory towards making these modifications and we intend to report on the progress and results in the near future.

## 8. Conclusions

To summarize, we have demonstrated an optically controlled waveplate at ~1323 nm using a ladder transition in a Rb vapor cell. We are able to place the probe in any arbitrary polarization by controlling the properties of the pump such as pump power, detuning and polarization. The process is explained in terms of the selection rules for the Zeeman sublevels. The waveplate has applications in Stokesmetric Imaging and optical switching. Using a comprehensive model which incorporates all the relevant Zeeman sub-levels in the system, we identify the parameters for optimal performance. A novel algorithm to compute the evolution of large scale quantum system enabled us to perform this computation. When combined with an optically controlled polarizer, recently demonstrated by us, and using a tapered nanofiber system, such a waveplate can be used to realize a Quantum Zeno Effect based ultra-low power, all-optical logic gate for the telecom band. We have identifed modifications to the system that are necessary in order to achieve a greater phase shift and also discussed some issues related to implementation in the TNF system.

This work was supported in part by AFOSR Grant # FA9550-10-01-0228, and the DARPA ZOE program under Grant # W31P4Q-09-1-0014.